\documentclass[twocolumn,showpacs,prb,amsmath,amssymb,floatfix]{revtex4}
\usepackage{amsmath,amssymb}
\usepackage{bm}
\usepackage{epsfig}

\begin{document}
  \title{Linear Magnetic Response of Disordered Metallic Rings: Large Contribution
from Forward Scattering Interactions}
  
  \author{Long Phi  Chau and Peter Kopietz}
  
  \affiliation{Institut f\"{u}r Theoretische Physik, Universit\"{a}t
    Frankfurt, Robert-Mayer-Strasse 8, 60054 Frankfurt, Germany}
   
  \date{February 12, 2004}

  \begin{abstract}
We calculate the effect of electron-electron interactions 
involving vanishing momentum transfer (forward scattering) on the
orbital linear magnetic response of
disordered metal rings pierced by a magnetic flux $\phi$.
Using the bulk value of the Landau parameter $F_0$ for copper, 
we find that
in the experiment by L\'{e}vy {\it{et al.}}
[Phys. Rev. Lett. {\bf{64}}, 2074 (1990)] the 
forward scattering contribution to the {\it{linear}}  magnetic response
is larger than the corresponding contribution from
large momentum transfers considered by 
Ambegaokar and Eckern   [Phys. Rev. Lett. {\bf{65}}, 381 (1990)].
However, outside the regime of validity of linear response 
and to first order in the  effective screened  interaction
the persistent current is dominated by
scattering processes involving large momentum transfers.

  \end{abstract}




  \maketitle

\section{Introduction}

\begin{figure}[tb]
    \begin{center}
     \epsfig{file=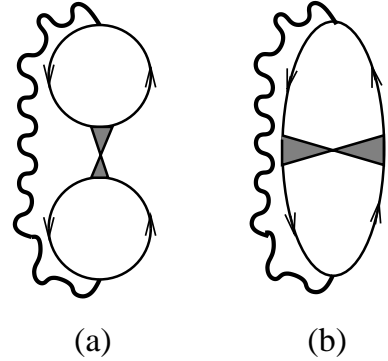,width=50mm}
    \end{center}
    \vspace{-4mm}
    \caption{%
Feynman  diagrams representing    
the flux-dependent part of the
    grand canonical potential  to 
 first order in the screened interaction.
(a) Hartree diagram; (b) Fock diagram.
Solid arrows represent non-interacting disorder averaged Green functions and
thick wavy lines represent the effective density-density interaction.
The Cooperon (shaded symbol) is defined in Fig.~\ref{fig:CD}.}
    \label{fig:AE}
  \end{figure}

More than a decade ago
the measurement by L\'{e}vy {\it{et al.}} \cite{Levy90} of 
persistent currents in mesoscopic  normal metal rings
pierced by an Aharonov-Bohm flux $\phi$  has
triggered a lot of theoretical activity \cite{Imry97,Schwab02}. 
Yet, up until now a truely convincing and generally accepted
theoretical explanation of the surprisingly large
persistent currents observed in Ref.  \cite{Levy90} and in subsequent
experiments \cite{Mohanty99,Jariwala01} has not been found.
It has become clear, however, that this effect cannot be explained
within a model of non-interacting electrons.
Ambegaokar and Eckern (AE) \cite{Ambegaokar90} were the first to
examine the effect of  electron-electron interactions on  
mesoscopic persistent currents: they realized that,
 to first order in the screened 
Coulomb-interaction,
 the dominant contribution to the
disorder averaged persistent current can be obtained from the two
diagrams shown in Fig. \ref{fig:AE},  representing a special
correction  $  \overline{\Omega}_{\rm AE} ( \phi )$ 
to the disorder averaged 
thermodynamic potential
which depends strongly on the Aharonov-Bohm flux $\phi$.
  \begin{figure}[tb]
    \begin{center}
     \epsfig{file=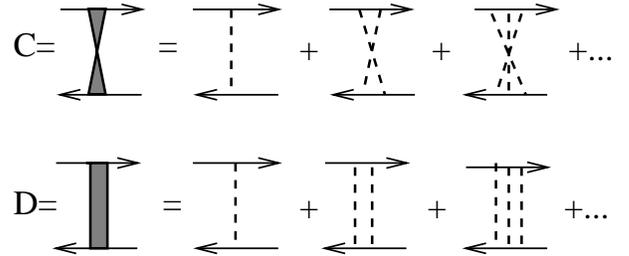,width=80mm}
    \end{center}
    \vspace{-4mm}
    \caption{%
Diagrammatic definitions of the Cooperon (C) and the Diffuson (D).
A dashed line represents the
covariance of the impurity potential.
}
    \label{fig:CD}
  \end{figure}
Here  the overline denotes averaging
over the disorder.
Given the grand canonical potential $\Omega ( \phi )$,
the corresponding 
persistent current $I ( \phi )$ can be obtained
from  the thermodynamic relation
 \begin{equation}
  I ( \phi )   = - c \frac{\partial \Omega ( \phi )  }{ \partial \phi }
 \; .
 \end{equation}

In a bulk metal  at high densities the
bare Coulomb-interaction 
$V_{0} ( {\bf{q}} ) = 4 \pi e^2 / {\bf{q}}^2$ is
strongly screened. A simple way to take the screening into account 
diagrammatically is the random-phase approximation (RPA).
Following this procedure,
AE approximated the effective interaction
(in the imaginary frequency formalism) as follows  
 \begin{equation}
 \overline{V}_{\rm RPA} ( {\bf{q}} , i \omega ) = \frac{ V_0 ( {\bf{q}} ) }{
1 +  \overline{ \Pi}_0 ( {\bf{q}} , i \omega ) V_0 ( {\bf{q}} ) }
 \; .
 \label{eq:RPA}
 \end{equation} 
For  momentum transfers $| {\bf{q}}|$ small compared
with the inverse elastic mean free path $\ell^{-1}$, 
and for frequency transfers $| \omega |$ small compared with the
inverse elastic lifetime $\tau^{-1}$
the disorder averaged polarization
is given by
 \begin{equation}
 \overline{ \Pi}_0 ( {\bf{q}} , i \omega ) 
 \approx 2 \nu_0 \frac{ D_0 {\bf{q}}^2}{ 
 D_0 {\bf{q}}^2 + | \omega |  }
\; ,
 \label{eq:pol}
 \end{equation} 
where $D_0$ is the diffusion coefficient
and $\nu_0$ is the average density of states 
at the Fermi energy (per spin) in the absence of interactions.
Note that $\nu_0 = ( \Delta_0 {\cal{V}} )^{-1}$, where ${\cal{V}}$ is the volume
of the system and $\Delta_0$ is the average level spacing (per spin)
at the Fermi energy.
It turns out that both diagrams in Fig. \ref{fig:AE} are dominated
by momentum transfers of the order of the Fermi momentum
$k_F$, which for a metallic system is
large compared with $\ell^{-1}$.
Eqs.~(\ref{eq:RPA}) and (\ref{eq:pol}) are therefore not suitable
for a  quantitatively accurate calculation of  persistent currents.
To make some progress analytically, AE estimated the
contribution from the diagrams in Fig. \ref{fig:AE}
by replacing the
effective interaction by a constant
 \begin{equation}
  \overline{V}_{\rm RPA} ( 
  {\bf{k}} - {\bf{k}}^{\prime} , i \omega ) 
 \rightarrow
  \langle \overline{V}_{\rm RPA} ( 
  {\bf{k}}_F - {\bf{k}}_F^{\prime} , i 0 ) \rangle
 \equiv \overline{V}
 \; ,
 \label{eq:Vbar}
 \end{equation}
where $\langle \ldots \rangle $ denotes the Fermi surface average over 
${\bf{k}}_F$ and ${\bf{k}}_F^{\prime}$.
For simplicity, it is assumed that the ring is quasi one-dimensional,
with transverse thickness $L_{\bot}$ in the range
$ k_F^{-1} \ll L_{\bot} \ll \ell \ll L$, where $L$ is the circumference of the ring. 
Then diffusive motion is only possible along the circumference.
At temperature $T=0$ the resulting average persistent current can be 
written as \cite{Ambegaokar90}
 \begin{equation}
 \overline{I}^{\rm AE} ( \phi ) = \sum_{k=1}^{\infty}
 I_k^{\rm AE} \sin ( 4 \pi k \phi / \phi_0 )
 \; ,
 \label{eq:IAE}
 \end{equation}
where $\phi_0 = hc / e$ is the flux quantum \cite{weprefer} 
and the Fourier coefficients of the current are
 \begin{equation}
 I_k^{\rm AE}
 = \frac{c}{\phi_0}  \frac{16 \lambda_c  }{k^2} E_c
 e^{ - k \sqrt{\gamma}} [ 1 + k \sqrt{{\gamma}} ]
 \; .
 \label{eq:ImAE}
 \end{equation}
Here $E_c =  \hbar D_0 / L^2$ is the Thouless energy and 
$\gamma =  \Gamma / E_c \ll 1$,  where 
at zero temperature $\Gamma = \Delta_0 / \pi$ 
is the cutoff energy
that regularizes the singularity in the Cooperon in a finite system~\cite{Voelker96}, see
Eqs. (\ref{eq:gwl}) and (\ref{eq:Pdef}) below.
The coupling constant ${\lambda}_c  =   \nu_0 \bar{V} $ can be identified with the 
dimensionless effective interaction in the Cooper channel to first order
in perturbation theory.
AE estimated
 ${\lambda}_{c} \approx 0.3 $, assuming
that the validity of the RPA can be extended to momentum transfers 
of the order of $k_F$. However, 
higher order ladder  diagrams in the Cooper channel strongly reduce 
the effective interaction, so that $\lambda_c \approx 0.06$ is a 
more realistic estimate \cite{Eckern91} for the Cu-rings in the experiment~\cite{Levy90}.

In real space Eq. (\ref{eq:Vbar}) amounts to
replacing the electron-electron interaction by a local effective density-density interaction,
 \begin{equation}
 \overline{V}_{\rm eff} ( {\bf{r}} - {\bf{r}}^{\prime} ) \rightarrow
  \overline{V} \delta ( {\bf{r}} - {\bf{r}}^{\prime} )
 \; .
 \label{eq:Veff}
 \end{equation}
More precisely, this replacement means that
for distances $| {\bf{r}} - {\bf{r}}^{\prime} |$ larger than  $ \ell$,
the interaction is effectively local.
In a recent letter Schechter, Oreg, Imry, and
Levinson \cite{Schechter03} 
pointed out that a different type of effective interaction
can possibly lead to a much larger persistent current.
Specifically, they used the BCS model to calculate
the leading interaction correction to
the orbital linear magnetic response and found \cite{Schechter03,weprefer}
 \begin{equation}
   \left. \frac{\partial \overline{  I }^{\rm BCS} }{\partial \phi } 
  \right|_{\phi =0} 
 = \frac{c}{\phi_0^2}    32 \pi \lambda_{\rm BCS} E_c \ln \left( \frac{ E_{\rm co}}{
 \Delta_0 } \right)
 \; ,
 \label{eq:chiBCS}
 \end{equation} 
where $\lambda_{\rm BCS}<0 $ is the attractive dimensionless interaction in the BCS model, 
and the coherence energy $E_{\rm co}$ is the smaller energy of
$\hbar / \tau$ and the Debye energy $ \hbar \omega_D$.
Eq. (\ref{eq:chiBCS}) should be compared with the
corresponding result for the local interaction model
used by AE, which implies according to Eqs. (\ref{eq:IAE}) and (\ref{eq:ImAE}),
 \begin{equation}
   \left. \frac{\partial \overline{  I }^{\rm AE} }{\partial \phi } 
  \right|_{\phi =0} 
 = \frac{c}{\phi_0^2}   32 \pi \lambda_{c} E_c \ln \left( \frac{ E_{c}}{
 \Delta_0 } \right)
 \; ,
 \label{eq:chiAE}
 \end{equation} 
where we have used $\Gamma = \Delta_0 / \pi $ and
retained only the leading logarithmic order. 
Note that the logarithm
is due to the slow decay ($ \propto k^{-1} $) of the 
Fourier coefficients $4 \pi k I_k^{\rm AE} / \phi_0$ of 
 $ \partial \overline{  I }^{\rm AE}   / {\partial \phi }$, 
so that all coefficients with  $ k \lesssim 1/\sqrt{\gamma}$
contribute to the linear response.
For $E_{ \rm co} \gg E_c$ the linear magnetic response in the
BCS model is parametrically larger than the linear response
in the local interaction model.
Whether or not this remains true beyond the linear response has not been 
clarified.
Note also that 
in the BCS model  the linear magnetic response 
is diamagnetic because  the effective interaction is attractive ($\lambda_{\rm BCS} < 0$), 
whereas 
the linear response in the local interaction model is
paramagnetic, corresponding to a repulsive effective interaction  ($\lambda_c > 0)$.

\section{Magnetic response due to forward scattering}

\begin{figure}[tb]
    \begin{center}
     \epsfig{file=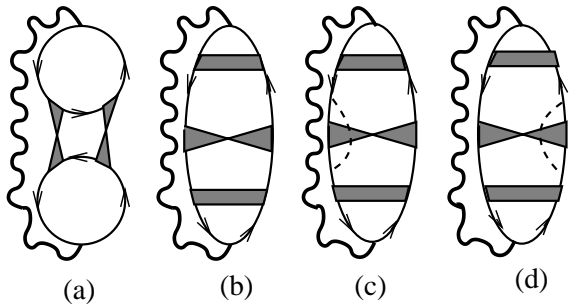,width=75mm}
    \end{center}
    \vspace{-4mm}
    \caption{%
    Feynman diagrams that dominate the
    flux-dependent part of the
    grand canonical potential 
 if the effective interaction involves only momentum transfers
smaller than the inverse elastic mean free path,  $  | {\bf{q}}  | \lesssim \ell^{-1}$.
 For vanishing momentum transfer the
   Hartree diagram (a) dominates the linear magnetic response,
    whereas outside the regime of validity of linear response
   the sum of the three Fock diagrams (b)--(d) has 
   the same order of magnitude as the Hartree diagram (a).
   Note that the Diffuson (shaded box, see Fig.~\ref{fig:CD}) 
renormalizes only the
density vertex in the  Fock diagrams; the
interaction in the Hartree diagram does not transfer any energy and hence
cannot be renormalized by singular Diffuson corrections.
  }
    \label{fig:AS}
  \end{figure}

An interesting observation made by the authors of Ref.~\cite{Schechter03}
is that an effective interaction different from the local interaction
used by AE can lead to a much larger persistent current, at least for sufficiently small
flux $\phi$, where it is allowed to calculate the current from the linear response.
Given the rather crude approximations in the microscopic derivation of the
local interaction model, it seems worth while to explore
the magnetic response for 
other types of effective interactions.
A possibility  which  so far has not been thoroughly 
analyzed  is an interaction which
is dominated by small momentum transfers.
Note that the  assumption that 
only forward scattering
processes    (corresponding to vanishing momentum transfer)
have to be taken into account for a consistent description of 
the low-energy and long-wavelength properties of normal
metals lies at the heart of the Landau's Fermi liquid theory. 
The  Landau model is in a sense the opposite extreme of the
local interaction model, because the effective interaction in the Landau model
is proportional to a Kronecker-delta  in momentum space,
 \begin{equation}
 \overline{V}_{\rm eff} ( {\bf{q}} , i \omega )
 \rightarrow \delta_{ {\bf{q}} , 0 } f_0
 \; ,
 \label{eq:f0def}
 \end{equation}
where the Landau parameter $f_0$ can be determined from experiments.
In fact, the dimensionless Landau parameter~\cite{notethat}
$ F_0 \equiv  2 \nu_0 f_0 $ can be written as
 $F_0 = \frac{B}{B_0} \frac{ m_{\ast}}{m_0}  -1$,
where $B$ is the bulk modulus, $m_{\ast}$ is the effective mass, and
$B_0$ and $m_0$ are the corresponding quantities in the absence 
of interactions.  Inserting the known bulk values for Cu \cite{Ashcroft76},
$m_{\ast}/ m \approx 1.3$ and $B/B_0 \approx 2.1$, we find
$F_0 \approx 1.7$, which is a factor of $30$ larger than the corresponding
estimate $\lambda_c \approx 0.06 $ in the local interaction model.
Note that in real space Eq. (\ref{eq:f0def}) corresponds to a constant 
effective interaction, proportional to the inverse volume of the system
 \begin{equation}
 \overline{V}_{\rm eff} ( {\bf{r}} - {\bf{r}}^{\prime} ) \rightarrow
  \frac{f_0}{\cal{V}}
 \; .
 \label{eq:Veff2}
 \end{equation}
Given an effective interaction of the form (\ref{eq:f0def}), 
the dominant flux-dependent contributions to the average
potential $\overline{\Omega} ( \phi )$ to first order in the interaction
are shown in Fig. \ref{fig:AS}.
The Fock diagrams (b)--(d) have been discussed previously in 
Refs.~\cite{BealMonod92,Kopietz98}; as first pointed out  by 
B\'{e}al-Monod and Montambaux \cite{BealMonod92},
to leading order in the small parameter $ (k_F \ell )^{-1}$, the  three Fock diagrams 
in Fig.~\ref{fig:AS} (b)--(d)
cancel, so that a direct  evaluation of the sum of these diagrams
is rather difficult. To calculate the leading contribution of these diagrams,
we note that the fermion loops in Fig.~\ref{fig:AS} (b)--(d) 
can be identified with contributions to the disorder averaged
polarization, which for general frequencies and small wavevectors
can be written as \cite{Vollhardt80}
 \begin{equation}
 \overline{ \Pi}_0 ( {\bf{q}} , i \omega ) 
 = 2 \nu_0 \frac{ D ( i \omega ) {\bf{q}}^2}{ 
 D ( i \omega ) {\bf{q}}^2 + | \omega |  }
\; ,
 \label{eq:pol2}
 \end{equation} 
where $D ( i \omega )$ is a generalized frequency-dependent Diffusion coefficient.
The crucial observation is now that the sum of the three Fock diagrams 
in Fig.~\ref{fig:AS} (b)--(d) corresponds to the usual weak localization
correction to the average conductance~\cite{Kopietz98},
 \begin{equation}
  D ( i \omega ) \approx  D_0 [ 1 + g_{\rm WL} ( i \omega ) ]
 \; ,
 \end{equation}
where
 \begin{equation}
 g_{\rm WL} ( i \omega ) = - \frac{2 \Delta_0}{\pi} \sum_{ {\bf{q}}} 
 \frac{1}{ \hbar D_0 {\bf{q}}^2 + | \omega | + \Gamma }
 \; .
 \label{eq:gwl}
 \end{equation}
Essentially we have used the equation of continuity to replace
the charge  vertices in Fig.~\ref{fig:AS} by current vertices,
which cannot be renormalized by singular diffusion 
corrections. The fact that a gauge transformation
replacing charge vertices by current vertices can be used to
avoid the explicit calculation of vertex corrections
has also been employed in Ref.\cite{Kopietz98PRL} to calculate the
zero bias anomaly in the tunneling density of states of
two-dimensional disordered electrons interacting with Coulomb forces. 

The evaluation of the contribution of the three Fock diagrams  in Fig.~\ref{fig:AS}
to the persistent current is now straightforward.
Note that for a thin ring with $ L_{\bot} \ll \ell \ll L$ the
${\bf{q}}$-summation  is one-dimensional, with quantized wave-vectors
$ 2 \pi ( n + 2 \phi/ \phi_0 ) /L$, $n = 0, \pm 1 , \pm 2 , \ldots$.  
Then we obtain
for the $k$-th Fourier component of the average current  due to the
Fock diagrams (b)--(d) in Fig.~\ref{fig:AS}  for the Landau model~\cite{Kopietz98},
 \begin{equation}
I_k^{\rm L, Fock} \propto k^{-1} \frac{f_0 }{ {\cal{V}}}
 \; .
 \label{eq:ILFock}
 \end{equation}
Due to the extra factor of inverse volume,
this  contribution is, for experimentally relevant
parameters \cite{Levy90}, negligible compared with 
corresponding result
in the local interaction model given in Eq. (\ref{eq:ImAE}).

The Hartree diagram in the Landau model is more interesting.
The fact that the diagram with two  Cooperons
shown in Fig.~\ref{fig:AS} (a) dominates the persistent current
due to electron-electron interactions with momentum transfers
 $| {\bf{q}} | \lesssim \ell^{-1}$ 
has already been pointed out 
in Ref.~\cite{Kopietz93}.
A similar diagram with two Cooperons (but without  interaction line) 
dominates the  fluctuations of the number of energy levels
in a fixed energy window centered at the Fermi energy \cite{Altshuler86}.
Using the approximate relation 
 \begin{equation}
I_{N} ( \phi )  = - \frac{c}{2} {\Delta}_0
 \frac{
\partial (\delta N )^2 }{ \partial \phi}
 \end{equation}
 between the persistent current 
$I_N ( \phi )$ 
at constant particle number and
the fluctuation  $ ( \delta N )^2$ of the particle number at constant
chemical potential $\mu$, several authors have realized \cite{Schmid91,Altshuler91,Oppen91} that
without interactions
the two-Cooperon diagram  determines the average
persistent current in a canonical ensemble.
Note that the Hartree diagram 
in Fig.~\ref{fig:AS} (a) does not contain
any vertex corrections analogous to the
diffusion corrections of the vertices in the Fock diagams
(b)--(d).  This is due  to the fact that the interaction line
in the Hartree process does not transfer any energy. Hence, the two
Green functions attached to the vertex of a Hartree interaction are either both
retarded or both advanced, so that it is impossible to attach a singular
Diffuson  to the vertex.

For the Landau model 
the Hartree diagram in Fig.~\ref{fig:AS} (a) yields 
at finite temperature $T$ the following
correction to the disorder averaged grand canonical potential,
 \begin{eqnarray}
\overline{\Omega}^{\rm L, Hartree} ( \phi ) & = &  \frac{f_0}{2 \cal{V}}  4 \sum_{ {\bf{q}}} 
 T^2 \sum_{ \tilde{\omega}_n , \tilde{\omega}_{n^{\prime}} }
 \theta ( - \tilde{\omega}_n \tilde{\omega}_{n^{\prime}} )
 \nonumber
 \\
 & & \hspace{-15mm} \times  \left ( \frac{ \Delta_0  }{ 2 \pi } \frac{\hbar}{\tau} \right)^2
\left[ \frac{  \hbar / \tau  }{ 
  \hbar D_0 {\bf{q}}^2 + | \tilde{ \omega}_n - \tilde{\omega}_{n^{\prime}} |  + \Gamma } 
 \right]^2
 \nonumber 
 \\
 &  & \hspace{-15mm} \times
 \sum_{ {\bf{k}}} [ \overline{G}_0 ( {\bf{k}} , i \tilde{\omega}_n ) ]^2 \overline{G}_0 ( - {\bf{k}} + {\bf{q}} , 
  i \tilde{\omega}_{n^{\prime}} )
 \nonumber
 \\
 &  & \hspace{-15mm} \times
\sum_{ {\bf{k}}^{\prime}  } [ \overline{G}_0 ( {\bf{k}}^{\prime} , 
i \tilde{\omega}_{n^{\prime}} ) ]^2 \overline{G}_0 ( - {\bf{k}}^{\prime}  + {\bf{q}} , 
  i \tilde{\omega}_{n} )
 \; .
 \label{eq:OmegaH0}
 \end{eqnarray}
Here $ \tau = \ell / v_F$ is the elastic lifetime, $\tilde{\omega}_n = 2 \pi ( n + \frac{1}{2} ) T$ are
fermionic Matsubara frequencies, and  
 \begin{equation}
\overline{G}_0 ( {\bf{k}} , i \tilde{\omega}_n )
= \frac{1}{ i \tilde{\omega}_n - \frac{   \hbar^2 {\bf{k}}^2}{2m} + 
\mu + i \frac{\hbar}{2 \tau}  {\rm sign} \tilde{\omega}_n    }
 \end{equation}
is the disorder averaged non-interacting Matsubara Green function.
Since the  Cooperons (i.e. the second line) in Eq. (\ref{eq:OmegaH0}) 
are only singular for $ | {\bf{q}} | \lesssim \ell^{-1}$,
and because
the ${\bf{k}}$- and ${\bf{k}}^{\prime}$- sums
are dominated by momenta of the order of the Fermi momentum, we may approximate
$\overline{G}_0 ( - {\bf{k}} + {\bf{q}} , 
  i \tilde{\omega}_{n^{\prime}} )
\approx
\overline{G}_0 ( - {\bf{k}} , 
  i \tilde{\omega}_{n^{\prime}} )$ and
$
\overline{G}_0 ( - {\bf{k}}^{\prime}  + {\bf{q}} , 
  i \tilde{\omega}_{n} )
\approx
\overline{G}_0 ( - {\bf{k}}^{\prime}   , 
  i \tilde{\omega}_{n} )$
in Eq. (\ref{eq:OmegaH0}).
The product of the last two lines of Eq. (\ref{eq:OmegaH0}) gives then rise to a factor of
$ [ (2 \pi / \Delta_0 ) ( \tau / \hbar )^2 ]^2$, 
so that we obtain 
 \begin{equation}
 \overline{\Omega}^{\rm L, Hartree} ( \phi ) = \frac{f_0}{2 \cal{V}} P ( \phi )
 \; ,
 \label{eq:OmegaH}
 \end{equation}
with the dimensionless coefficient
 \begin{equation}
 P ( \phi ) = \frac{4 T}{\pi} 
 \sum_{ 0 < \omega_m < \hbar / \tau } \sum_{ {\bf{q}} }
 \frac{ \omega_m}{ [ \hbar D_0 {\bf{q}}^2 + \omega_m + \Gamma ]^2 }
 \label{eq:Pdef}
 \; ,
 \end{equation}
where $\omega_m = 2 \pi m T$ are bosonic Matsubara frequencies.
Assuming again a thin ring with $ L_{\bot} \ll \ell \ll L$, we find 
in the limit  $T \rightarrow 0$ for  the Fourier components of the
persistent current,
 \begin{equation}
 I_k^{\rm L, Hartree}
 =   \frac{16  }{\pi}   \frac{c}{\phi_0}  \frac{f_0}{ {2 \cal{V}}}
 e^{ - k \sqrt{\gamma}}
 =     \frac{c}{\phi_0} 8 F_0 \frac{ \Delta_0}{\pi}
 e^{ - k \sqrt{\gamma}}
 \; .
 \label{eq:IH}
 \end{equation}
 Comparing this expression with the
corresponding result (\ref{eq:ImAE}) of the local
interaction model, we see that in the Landau model
the Fourier components $I_k^{\rm L,Hartree}$ are independent of 
$k$ as long as $k \lesssim 1/ \sqrt{\gamma}$. 
Therefore the linear magnetic response
is determined by all Fourier components up to
$k \lesssim \sqrt{ E_c/ \Gamma}$,
 \begin{eqnarray}
 \left. \frac{\partial \overline{  I }^{\rm L, Hartree} }{\partial \phi } 
  \right|_{\phi =0} 
 & = &  \frac{ 4 \pi}{\phi_0} \sum_{k=1}^{\infty} k  I_k^{\rm L, Hartree}
 \nonumber
 \\
 &  \approx & \frac{c}{\phi_0^2}   16 \pi F_0 E_c 
\label{eq:chiL}
 \; ,
 \end{eqnarray}
where we have assumed that  $\gamma = \Gamma / E_c \ll 1$, so that
 \begin{equation}
 \sum_{k=1}^{\infty} k e^{- k \sqrt{\gamma }}
 = \frac{e^{\sqrt{\gamma}}}{[ 1 - e^{ - \sqrt{\gamma} }]^2}
 \approx \frac{1}{\gamma}
 \approx \frac{ \pi E_c}{\Delta_0}
 \; .
 \end{equation}
Note that the small energy scale $\Delta_0$ of Eq. (\ref{eq:IH}) has disappeared 
on the right-hand side of Eq. (\ref{eq:chiL}), and is replaced
by the much larger Thouless energy $E_c$.
Due to the faster decay of the Fourier components (\ref{eq:ILFock})
of the Fock contribution in the Landau model,
the linear response due to the Fock diagrams shown in Fig.~\ref{fig:AS} (b)--(d)
is a factor of $\sqrt{\Gamma / E_c }$ smaller than the
corresponding Hartree contribution.
Interestingly,  the anomalously large linear magnetic response in
the BCS model given in Eq. (\ref{eq:chiBCS}) is also dominated by the
Hartree process~\cite{Schechter03}. 
Thus, the importance of Hartree interactions
for persistent currents is some extent
independent of a specific model for the interaction. 
For the Cu-rings used in the experiment \cite{Levy90} we estimate
$F_0 \approx 1.7$, $\lambda_c \approx 0.06$ and $E_c / \Delta_0 \approx 25$;  with these values
the linear magnetic response due to  forward scattering is more than four times larger than the
linear response in the local 
interaction model considered by AE \cite{Ambegaokar90}. 
To take both contributions into account
one should parameterize the total effective interaction as
  \begin{equation}
 \overline{V}_{\rm eff} ( {\bf{r}} - {\bf{r}}^{\prime} ) =
  \overline{V} \delta ( {\bf{r}} - {\bf{r}}^{\prime} ) + 
\frac{f_0 - \overline{V}}{\cal{V}}
 \; ,
 \label{eq:Veff3}
 \end{equation}
which in momentum space amounts to 
   \begin{equation}
  \overline{V}_{\rm eff} (  {\bf{q}} ) = \left\{
 \begin{array}{cl}
  \overline{V} & \mbox{for $ {\bf{q}} = 0 $} \\
   f_0 &  \mbox{for $ {\bf{q}} \neq 0 $}
  \end{array}
  \right.
  \; .
  \end{equation}
Using the  estimate for bulk Cu given above,
$F_0 = 2 \nu_0 f_0 \approx 1.7$ and $ \lambda_c = \nu_0 \overline{V} \approx 0.06$, we find
$f_0 / \overline{V} \approx 14$, which supports our assumption that
there is indeed a strong
enhancement of the effective interaction in the forward scattering channel.

\section{Conclusions}

In summary,
we have shown that a forward scattering
excess interaction, involving   a vanishing  momentum transfer,
yields the dominant contribution to the {\it{linear}} magnetic response
of mesoscopic metal rings for experimentally relevant
parameters~\cite{Levy90}.
On the other hand,  outside the linear response regime the 
persistent current is
dominated by the term $\overline{V}$ involving large momentum transfers, at least 
if we use the bulk estimates for $\overline{V}$ and $F_0$ for Cu.
However, for a mesoscopic disordered Cu-ring it is not obvious that
 the bulk estimates are reliable. 
Note also that in the bulk the normal Fermi liquid  is stable as long as $F_0 > -1 $, so that
for $0> F_0 > -1 $
the linear magnetic response in the normal state
can be diamagnetic, in spite of the fact that the effective coupling $\lambda_c$
in the Cooper channel is positive.
Moreover, in the vicinity of an $s$-wave Pomeranchuk instability  \cite{Pomeranchuk58,Murthy03}, 
where $F_0 < 0$ and
$ | 1 + F_0  | \ll 1$,    
one should replace $F_0$
by $ F_0 / (1 + F_0)$. In this case we predict 
a strongly enhanced diamagnetic linear response. In fact, 
the forward scattering channel might then
dominate the persistent current even beyond the linear order in the flux $\phi$.
To clarify this point,
a better microscopic theory of the effective electron-electron interaction
in mesoscopic disordered metals 
is necessary.
In particular,  a microscopic theory should properly treat the problem of screening in a finite system and
incorporate the breakdown of Fermi liquid theory  
in quasi one-dimensional disordered metals
at sufficiently low temperatures \cite{Altshuler85}.

We thank M. Schechter for his clarifying remarks concerning
Ref. \cite{Schechter03}  and for his
comments on this manuscript.




\begin{thebibliography}{99}


\bibitem{Levy90}%
L.~P.\ L\'{e}vy, G.\ Dolan, J.\ Dunsmuir, and H.\ Bouchiat, Phys.\ 
Rev.\ Lett.\ {\bf{64}}, 2074 (1990).

\bibitem{Imry97}%
Y.\ Imry, {\it{Introduction to Mesoscopic Physics}} (Oxford
University Press, Oxford, 1997).

\bibitem{Schwab02}
U. Eckern and P. Schwab,  J. Low Temp. Phys. {\bf{126}}, 1291 (2002).

\bibitem{Mohanty99}
P. Mohanty, Ann. Physik (Leipzig) {\bf{8}}, 549 (1999).

\bibitem{Jariwala01}
E. M. Q. Jariwala, P. Mohanty, M. B. Ketchen, and R. A. Webb,
Phys. Rev. Lett. {\bf{86}}, 1594 (2001).

\bibitem{Ambegaokar90}%
V. Ambegaokar and U. Eckern, Phys. Rev. Lett. {\bf{65}}, 381 (1990).

\bibitem{weprefer}
We prefer to express the response in terms of the
normal flux quantum $\phi_0 = hc/e$, whereas 
$\Phi_0 = hc/2e$ in Ref. \cite{Schechter03} is the 
superconducting flux quantum. This is the reason why the numerical prefactor in 
our Eq. (\ref{eq:chiBCS}) is $32 \pi$, while the corresponding prefactor in Eq. (1) of
Ref. \cite{Schechter03} is $8 \pi$.

\bibitem{Voelker96}
A. V\"{o}lker and P. Kopietz,  Mod. Phys. Lett. B {\bf{10}}, 1397 (1996).

\bibitem{Eckern91}%
U. Eckern, Z. Phys. B {\bf{82}}, 393 (1991).

\bibitem{Schechter03}%
M. Schechter, Y. Oreg, Y. Imry, and Y. Levinson,
Phys. Rev. Lett. {\bf{90}}, 026805 (2003).
 
\bibitem{notethat}
Note that we define $\nu_0$ to be the density of states {\it{per spin}}, so that
$F_0 = 2 \nu_0 f_0$ agrees with the usual definition of the dimensionless Landau parameter.
%
\bibitem{Ashcroft76}
N. W. Ashcroft and N. D. Mermin, {\it{Solid State Physics}},  (Holt-Saunders, Philadelphia, 1976).

\bibitem{BealMonod92}
M. T. B\'{e}al-Monod and G. Montambaux, Phys. Rev. B {\bf{46}}, 7182 (1992).
%
\bibitem{Kopietz98}%
P. Kopietz and A. V\"{o}lker, Phys. Lett. A {\bf{244}}, 569 (1998).

\bibitem{Vollhardt80}
D. Vollhardt and P. W\"{o}lfle,
Phys. Rev. B {\bf{22}}, 4666 (1980).
%
\bibitem{Kopietz98PRL}
P. Kopietz, Phys. Rev. Lett. {\bf{81}}, 2120 (1998). 
%
\bibitem{Kopietz93}%
P. Kopietz, Phys. Rev. Lett. {\bf{70}}, 3123 (1993); erratum: {\bf{71}}, 306
(1993).
%
\bibitem{Altshuler86}
B. L. Altshuler and B. I. Shklovskii, Zh. Eksp. Teor. Fiz. {\bf{91}},
220 (1986) [Sov. Phys. JETP {\bf{64}}, 127 (1986)].
%
\bibitem{Schmid91}
A. Schmid, Phys. Rev. Lett. {\bf{66}}, 80 (1991).
%
\bibitem{Oppen91}
F. von Oppen and E. K. Riedel,
Phys.  Rev. Lett. {\bf{66}}, 84 (1991).
%
\bibitem{Altshuler91}
B. L. Altshuler, Y. Gefen, and Y. Imry, Phys. Rev. Lett. {\bf{66}}, 88 (1991).
%
\bibitem{Pomeranchuk58}
I. J. Pomeranchuk, 
Zh. Eksp. Teor. Fiz.  {\bf{35}}, 524 (1958)
[Sov. Phys. JETP {\bf{8}}, 361 (1958)].

\bibitem{Murthy03}
G. Murthy and R. Shankar, Phys. Rev. Lett. {\bf{90}}, 066801 (2003);
D. Herman, H.  Mathur, and G. Murthy,  Phys. Rev. B {\bf{69}}, 041301 (2004);
G. Murthy, R. Shankar, D. Herman, and H. Mathur, {\it{ibid.}} {\bf{69}}, 075321 (2004).
%
\bibitem{Altshuler85}
B. L. Altshuler and A. G. Aronov, in {\it{Electron-Electron Interactions
in Disordered Systems}}, edited by A. L. Efros and M. Pollak, (North Holland, Amsterdam, 1985).
%
%
%
%
%

 
  \end{thebibliography}
\end{document}